\documentclass[]{aastex}
\newcommand{\myemail}{itoh@hep01.hepl.hiroshima-u.ac.jp}

\begin{document}

\title{Minute-Scale Rapid Variability of Optical Polarization in
Narrow-Line Seyfert 1 Galaxy: PMN J0948+0022}

\author{
Ryosuke Itoh\altaffilmark{1},
Yasuyuki T. Tanaka\altaffilmark{1},
Yasushi Fukazawa\altaffilmark{1},
Koji S. Kawabata\altaffilmark{2},
Kenji Kawaguchi\altaffilmark{1},
Yuki Moritani\altaffilmark{2},
Katsutoshi Takaki\altaffilmark{1},
Issei Ueno\altaffilmark{1},
Makoto Uemura\altaffilmark{2},
Hiroshi Akitaya\altaffilmark{2},
Michitoshi Yoshida\altaffilmark{2},
Takashi Ohsugi\altaffilmark{2},
Hidekazu Hanayama\altaffilmark{3},
Takeshi Miyaji\altaffilmark{3}
and
Nobuyuki Kawai\altaffilmark{4} 
}

\altaffiltext{1}{Department of Physical Sciences, Hiroshima
University, Higashi-Hiroshima, Hiroshima 739-8526, Japan; \myemail}
\altaffiltext{2}{Hiroshima Astrophysical Science Center, Hiroshima
University, Higashi-Hiroshima, Hiroshima 739-8526, Japan}
\altaffiltext{3}{Ishigakijima Astronomical Observatory,
National Astronomical Observatory of Japan, 1024-1 Arakawa, Ishigaki, Okinawa 907-0024, Japan}
\altaffiltext{4}{Department of Physics, Tokyo Institute of Technology, 2-12-1 Ookayama, Meguro-ku, Tokyo 152-8551, Japan}

\begin{abstract}

We report on optical photopolarimetric results of the radio-loud
narrow line Seyfert 1 (RL-NLSy1) galaxy PMN J0948+0022 on 2012
December to 2013 February triggered by flux enhancements in near
infrared and $\gamma$-ray bands.
Thanks to one-shot polarimetry of the HOWPol installed to the Kanata
telescope, we have detected very rapid variability in the
polarized-flux light curve on MJD 56281 (2012 December 20). 
The rise and decay times were about 140 sec and 180 sec, respectively.
The polarization degree (PD) reached $36 \pm 3$\% at the peak of the
short-duration pulse, while polarization angle (PA) remained almost
constant. 
In addition, temporal profiles of the total flux and PD showed highly
variable but well correlated behavior and discrete correlation
function analysis revealed that no significant time lag of more than
10 min was present. 
The high PD and minute-scale variability in polarized flux provides 
a clear evidence of synchrotron radiation from a very compact emission
region of $\sim 10^{14}$ cm size with highly ordered magnetic field.
Such micro variability of polarization are also observed in several blazar jets,
but its complex relation between total flux and PD are explained by multi-zone model 
in several blazars.
The implied single emission region in PMN J0948+0022 might be reflecting a difference 
of jets between RL-NLSy1s and blazars.

\end{abstract}


\section{Introduction}

Narrow-line Seyfert 1 (NLSy1) galaxies are a class of active galactic
nuclei (AGNs).
It is widely recognized that NLSy1 posses a relatively light central
black hole (BH) of $\sim10^6-10^7M_{\odot}$ accreting at a very high
rate near the Eddington limit
\citep[e.g.,][]{2008ApJ...685..801Y}.
Hence, NLSy1 is considered to be a young AGN growing toward a 
super massive BH which is
believed to have a mass of $10^8-10^9M_{\odot}$.
Thus, studying a class of NLSy1 provides us with knowledge about BH evolution.
NLSy1s are usually radio-quiet and only 7\% of NLSy1 are radio-loud (RL) objects
\citep{2003ApJ...584..147Z,2006AJ....132..531K}.
Owing to the high sensitivity of {\it Fermi} LAT instrument, $\gamma$-ray
emission from some of these RL-NLSy1s are detected and the presence of a third
class of AGNs with relativistic jets are confirmed from
multi-wavelength observation \citep{2009ApJ...707L.142A} following blazars and radio galaxies.
These discoveries opened interesting questions on the unified model of AGNs
such as the development of the relativistic jet.
Combined with the fact that their broadband spectra are
quite similar to those of blazars,
it is thought that at least these NLSy1s have powerful
relativistic jets and they are directed toward us.
Finding evidence of blazar-like behavior in RL-NLSy1 would help us to
understand the evolution of relativistic jets.

Polarized radiation is one of the evidences of synchrotron origin of jet emissions.
Therefore, optical polarimetric observations provide a strong
tool to probe jet structures \citep[e.g., ][]{2008Natur.452..966M,2010Natur.463..919A}.
The radiation in the optical band is thought to be a 
superposition of synchrotron and disk emission 
from its spectral shape of the {\it Fermi}-detected NLSy1s 
which has enough multi-wavelength data.
However, a timescale of micro variability with photopolarimetric
observation in RL-NLSy1 has never been studied so far.
It can provide an important and interesting tool to probe of the size
and structure of the emission region in the jet and for distinguishing
different models responsible for the variability.
Several important and interesting results 
of micro variability for blazars are reported
\citep[e.g.,][]{1996ASPC..110...17M,2008ApJ...677..906F}.

PMN~J0948+0022 \citep[also known as 2FGL J0948.8+0020, R.A. = $09^h
48^m 57.3201^s$,
decl. = $+00\arcdeg\ 22\arcmin\ 25\arcsec.558$,
J2000, z=0.5846,][]{2012ApJS..199...31N,2002ApJS..141...13B}
is classified as a RL-NLSy1 and displays variable emission from
radio to $\gamma$-ray bands, suggesting the presence of the relativistic
jet \citep{2003ApJ...584..147Z,2009ApJ...707L.142A}.
The optical polarization variability with monthly-timescale were detected 
in the past observations \citep{2013arXiv1301.0657E}.
Additionally, a high polarization degree (PD) of 18.8\% 
was also reported in April 2009 \citep{2011PASJ...63..639I}.
Micro variability (timescale within a few hours) of optical flux was also
observed on PMN~J0948+0022 in 2009 \citep{2010ApJ...715L.113L},
and this result supports the fact that the object carries the relativistic
jet with a small viewing angle.
Hence, PMN J0948+0022 is a good target to study the difference of
the relativistic jet 
between blazars and RL-NLSy1s.

Recently, PMN~J0948+0022 showed an extreme activity in the
near-infrared and GeV $\gamma$-ray bands on December 2012
\citep{2012ATel.4659....1C,2013ATel.4694....1D}.
In this paper, we present results of high-temporal-density optical monitoring
observations of PMN~J0948+0022 just after the December 18th 2012 (MJD 56279) near-infrared flare.

\section{Observation}

We performed the {\it g'}, {\it V} and {\it R$_C$}-band photometric
observations of PMN~J0948+0022
from 2012 December 20 to 2013 February 20, using the HOWPol (Hiroshima One-shot Wide field Polarimeter)
installed to the 1.5m Kanata telescope located at the
Higashi-Hiroshima Observatory, Japan \citep{2008SPIE.7014E.151K} and
the MITSuME (Multicolor Imaging Telescopes for Survey and Monstrous Explosions) 
installed to the 1.05m Murikabushi telescope
located at the Ishigaki island, Japan.
Reductions of optical data were performed under the standard procedure
of CCD photometry.
We performed the aperture photometry using \verb|APPHOT| packaged in
\verb|PYRAF|
\footnote{PYRAF is a product of the Space Telescope Science Institute, which is
operated by AURA for NASA.
http://www.stsci.edu/institute/software\_hardware/pyraf},
and the differential photometry with a comparison star taken in the
same frame of PMN~J0948+0022.
The position of the comparison star is R.A. = $09^h 49^m 00.4^s$, decl. =
$+00\arcdeg\ 22\arcmin\ 35\arcsec.1$ (J2000)
and its magnitudes are  {\it g'} = 18.288 mag,{\it V} = 17.429 mag and {\it
R$_C$} = 16.849 mag \citep{2007ApJS..172..634A}.
We confirmed a systematical flux difference of the
comparison star between the two instruments and standard deviations of flux are $\sim 0.02$ mag in the
{\it V} band and $\sim 0.02$ mag in the {\it R$_C$} band.
These values were added to the photometric errors of PMN~J0948+0022 in
each observation.
We corrected the data for the Galactic extinction of  $A_{\rm g'}$=0.305 mag, 
$A_{\rm V}$=0.253 mag and  $A_{\rm R_{{\rm C}}}$=0.206 mag
\citep[][NED database\footnote{http://ned.ipac.caltech.edu/}]{2011ApJ...737..103S}.

We also performed temporally high-densed photopolarimetric observations
of PMN~J0948+0022 with the HOWPol on 6 nights.
In order to study a relation between the optical flux and
the micro variability,
we performed high-dense photopolarimetric observations in several
flux levels.
Thanks to the double-Wollaston prism installed to the HOWPol, we could
obtain both the Stokes $Q$ and $U$
parameters in only a single exposure.
Therefore short-interval (100-310 s) photopolarimetric observations
were available.
A detail photopolarimetric observation log is given in Table \ref{tab:obs}.
The bad quality frame (cloudy sky, suffering from cosmic ray and so on) are
excluded from our analysis.
Polarimetry with the HOWPol suffers from large instrumental polarization
($\Delta {\rm PD} \sim 4$\%)
caused by the reflection of the incident light on the tertiary
mirror of the telescope.
The instrumental polarization was modeled as a function of the
declination of the object and the
hour angle at the observation, and we subtracted it from the observed value.
The polarization angle (PA) is defined in the
standard manner as measured from north to east.
The PA was calibrated with two polarized stars, HD183143 and HD204827
\citep{1983A&A...121..158S}.
Because the PA has an ambiguity of $\pm 180^{\circ} \times n$ (where $n$ is an integer),
we selected $n$ which gives the least angle
difference from the previous data, assuming that
the PA would change smoothly.
We confirmed that the error of PD in the instrumental polarization
correction is smaller than 0.5\% and the error of PA is smaller 
than $2^{\circ}$ from observations of unpolarized and polarized stars.
The typical standard deviations of the Stokes parameters of the
comparison star during the observation were
$\Delta Q\sim 0.02$ and $\Delta U \sim 0.02$ which corresponds to
$\Delta {\rm PD} \sim 3\%$.
These values were added to the polarimetric errors of
PMN~J0948+0022 in each observation.

\begin{table}
  \centering
  \caption{Log of high dense photopolarimetric observations using HOWPol}
  \label{tab:obs}
  \begin{tabular}{lcrc}\hline\hline
    MJD   & Start \& End time (UT) & Interval$^*$ & $N_{\rm obs}$ \\ \hline
    56281 & 18.04 -- 20.94    & 160 s  & 51 \\
    56283 & 17.17 -- 20.69    & 220 s  & 26 \\
    56284 & 17.76 -- 20.18    & 260 s  & 22 \\
    56295 & 16.59 -- 18.86    & 310 s  & 14 \\
    56342 & 17.74 -- 19.21    & 100 s  & 16 \\
    56343 & 13.11 -- 15.88    & 160 s  & 18 \\
    \hline
  \end{tabular}
  \\ $^*$ CCD read out time (10 s) is included.
\end{table}

\section{Results}

\begin{figure}
  \centering
  \includegraphics[angle=0,width=15cm]{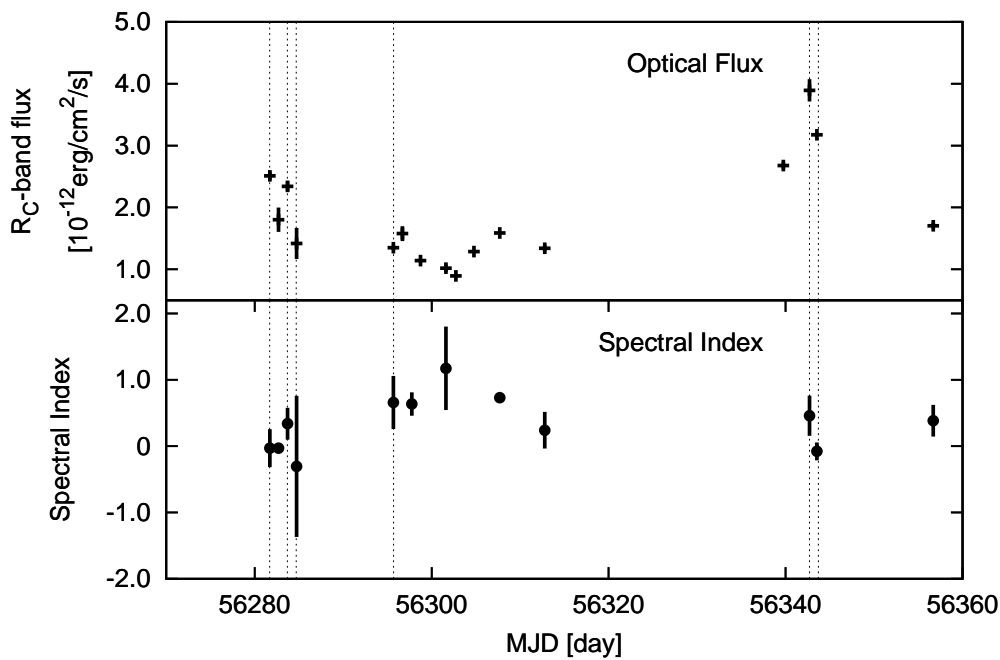}
  \caption{A long-term history of the optical flux and
    spectral index taken by the HOWPol and MITSuME. 
    Upper panel shows a light curve in the {\it R$_{\rm C}$} band. 
    Lower panel shows a history of spectral index.
    Dashed lines indicate the dates when high-dense photopolarimetric
    observation was performed.}
  \label{fig:LC_month}
\end{figure}

Figure \ref{fig:LC_month} shows a long-term history of {\it
R$_C$}-band flux and spectral index from
2012 December 20 (MJD 56281) to 2013 February 20 (MJD 56356).
A local two-point spectral index $\alpha$ is defined between two
brightness measurements $F_1$, $F_2$
at frequencies  $\nu_1$ and $\nu_2$, respectively, as
$\alpha = \ln (F_1/F_2)/\ln (\nu_1/\nu_2)$.
We use the mean wavelengths of the filters, 658.8 nm for the {\it
R$_{\rm C}$} band, 550.5 nm for the {\it V}
band and 485.8 nm for the {\it g'} band.
We calculate the spectral index with
the {\it R$_{\rm C}$} and the {\it V} band data except the data on MJD 56282,
on which the spectral index is calculated with the {\it R$_{\rm C}$} and the
{\it g'} band data.
The error bars include both statistical and systematic ones.
Large variation has been detected in the {\it R$_{\rm C}$}-band flux
and an amplitude of total flux reaches a factor of 4.
On the other hand, the variation of the spectral index is relatively
small and there are no clear
correlation between the optical flux and the spectral index.

\begin{figure*}
  \centering
  \includegraphics[angle=0,width=15cm]{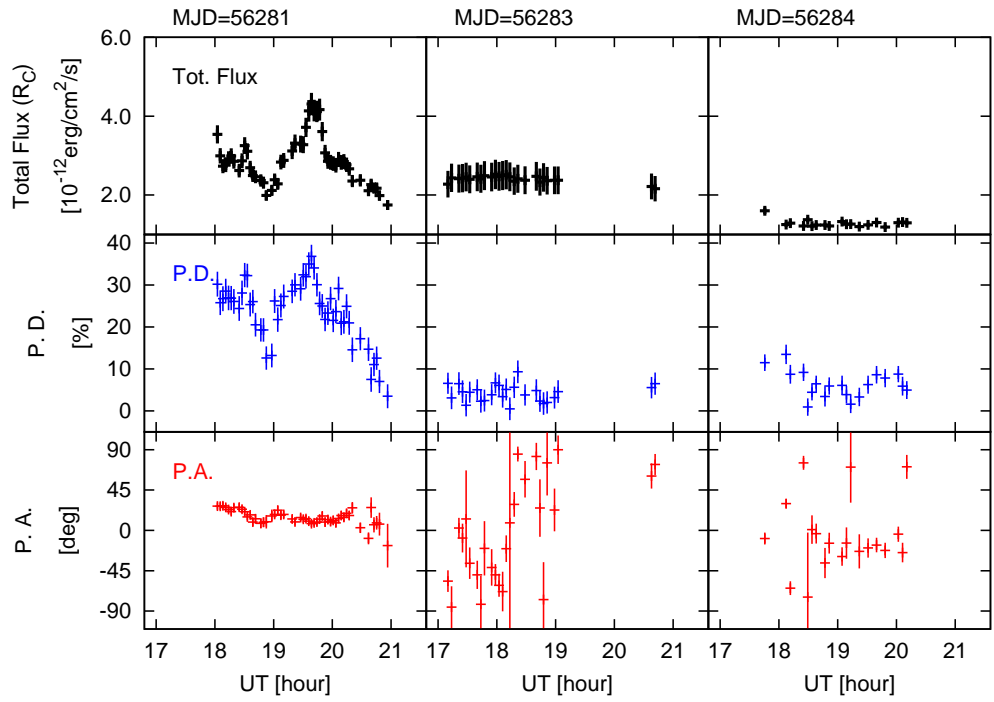}
  \caption{Intra-night light curve of PMN~J0948+0022 from MJD 56281 to 56284.
    From top to bottom, histories of  total flux in the {\it R$_{\rm C}$} band,
    polarization degree (PD) and  
    polarization angle (PA) are shown.}
  \label{fig:LC_daily1}
\end{figure*}

\begin{figure*}
  \centering
  \includegraphics[angle=0,width=15cm]{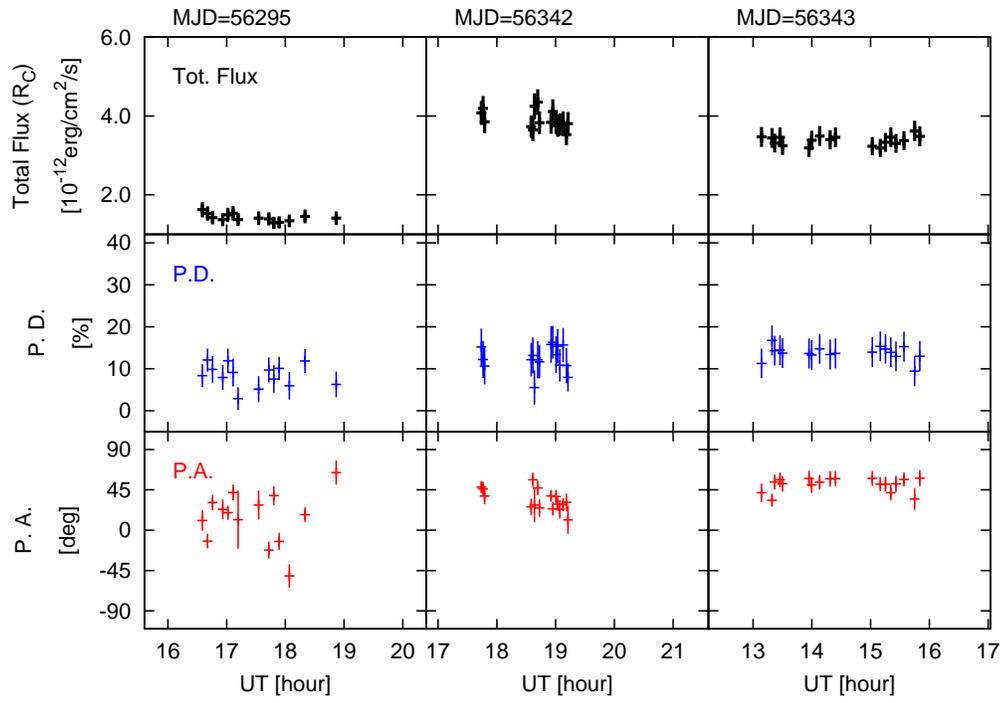}
  \caption{Intra-night light curve of PMN~J0948+0022 from MJD 56295 to 56343.
           The captions are same as in Figure~\ref{fig:LC_daily1}.}
  \label{fig:LC_daily2}
\end{figure*}

\begin{figure}
  \centering
  \includegraphics[angle=0,width=15cm]{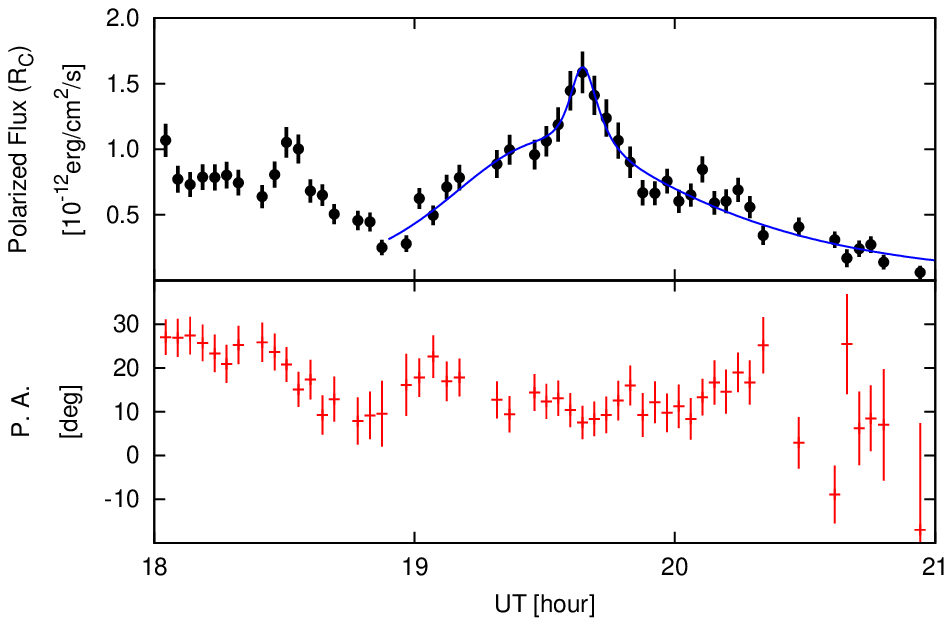}
  \caption{History of the polarized flux
    and polarization angle on MJD 56281.
    Solid line is the best-fit light curve. See the detail on the text.
   }
  \label{fig:MJD56281}
\end{figure}

Figure \ref{fig:LC_daily1} and \ref{fig:LC_daily2} show results of temporally high-dense
monitoring of PMN~J0948+0022
and figure \ref{fig:MJD56281} shows an enlarged view of temporal
variation of the polarized flux and PA on MJD 56281.
Light curves of $R_{\rm C}$-band polarized flux (PF) are calculated by
$PF = F_{\rm R_{\rm C}}  \times  PD/100$,
where $PD$ and $F_{\rm R_{\rm C}}$ are a measured polarization
degree in unit of \% and a total flux in the $R_{\rm C}$ band.
Rapid and violent outburst of optical flux and PF was clearly detected
on MJD 56281.
The object brightened by a factor of 2 in the total flux and by a factor
of 6 in the PF around 19.6 UT
within a few hours.
The variation pattern of PD is quite similar to that of total flux.
Discrete correlation function \citep[DCF;][]{1988ApJ...333..646E}
between the {\it R$_{\rm C}$}-band total flux and PD on MJD 56281
shows a good correlation (DCF value of 0.92) with no significant time
lag; an upper limit of lag is $\pm10$ minutes.
In the outburst state, PD reaches $36 \pm 3\%$ at maximum on MJD 56281.
This value is twice as high as that in the past observations of this object
\citep{2011PASJ...63..639I}.

The light curve on MJD 56281 has some structures.
It seems that there are two flaring components which have a different
variability timescale.
We tried a light curve fitting with the following function to reproduce the time
profile of outburst around 19.6 UT \citep{2010ApJ...722..520A}:
\begin{equation}
{\rm PF}(t) = {\rm PF}_C + {\rm PF}_{{\rm long}} \left( e^{\frac{t_{{\rm
long}}-t}{T_{r{{\rm long}}}}} + e^{\frac{t-t_{{\rm long}}}{T_{d{{\rm
long}}}}} \right)^{-1}
+ {\rm PF}_{{\rm short}} \left( e^{\frac{t_{{\rm short}}-t}{T_{r{{\rm
short}}}}} + e^{\frac{t-t_{{\rm short}}}{T_{d{{\rm short}}}}}
\right)^{-1},
\end{equation}
where ${\rm PF}_C$ represents an assumed constant level underlying the
outburst, ${\rm PF}_{{\rm long}}$ and ${\rm PF}_{{\rm short}}$ measures the
amplitude of the flares, $t_{{\rm long}}$ and $t_{{\rm short}}$
describe approximately the time of the peak,
$T_{r{{\rm long}}}$ and $T_{r{{\rm short}}}$ measure the rise time and
$T_{d{{\rm long}}}$ and $T_{d{{\rm short}}}$ measure the decay time.
Suffixes mean the long-term and the short-term component.
We obtained
$T_{r{{\rm long}}}  = 16 \pm 3$     min and $T_{d{{\rm long}}}  = 39
\pm 4$    min for the long-term component and
$T_{r{{\rm short}}} = 2.4 \pm 1.5$  min and $T_{d{{\rm short}}} = 3.0
\pm 1.6$ min for the short-term component.
The characteristic flare duration can be estimated as $T_r + T_d$.
There is a significant difference of variability timescale between two flares.
In contrast, PA maintains its value during the outburst around 19.6 UT
with an average value of $15 \pm 1$ deg.
We also obtained the upper limit of ${\rm PF}_C < 0.2 \times 10^{-12}$ erg cm$^{-2}$ s$^{-1}$.

Except the MJD 56281, there is no significant micro variability of
total flux and PD while the
daily light curve indicates significant variability from MJD 56283 to 56284.
Data on MJD 56283, 56284 and 56295 show a low PD value of $\sim 5\%$.
It seems that there is micro variability of PAs during these periods, but it
could be due to low statistics.
In MJD 56342 and 56343, 
a relatively increased total flux was observed with
no significant micro variability.
During these periods, a total flux is constant but higher than that of MJD 56281.
There are also no significant micro variability of PAs on MJD 56342 and 56343
but a gradual increasing of PA is implied.
PD is also higher than that on MJD 56283 and 56284.
The average of PAs is $36 \pm 3$ and $51 \pm 2$ deg in MJD 56342 and
56343, respectively, and these are slightly different
from that on MJD 56281 (PA is $15 \pm 1$ deg).

\section{Discussion}

Thanks to one-shot photopolarimetry of the HOWPol, we obtained the
highly variable light curve of polarized flux on MJD 56281.
In particular, the light curve in the latter night of the same day
(around 19--21 UT) consisted of two major pulses whose variability
timescales (rise and decay times) are of the order of minute and hour. 
No significant time lags between flux and PD would indicate that the 
short-duration pulse occurred in the shock-compressed region with highly ordered magnetic field.
This minute-scale variability suggests a very compact emission region
of the order of $R \sim c \delta t_{\rm var} = 1 \times 10^{14} \left(
\delta /10 \right) \left( t_{\rm var}/320 \ {\rm s} \right)$ cm, 
where $R$ is typical size of the emission region, $c$ is the speed of
light, $\delta$ is beaming factor, and $t_{\rm var}$ is the observed
variability timescale. 
In addition, the very high PD of up to $\sim$36\% observed during the
short-duration pulse requires highly ordered magnetic field inside the
very compact emission region.

NLSy1s are in general believed to have relatively smaller 
central BH mass compared to typical blazars (see \S1). 
If we assume that the variability timescale is determined by the activity of central BH, 
its lower limit is the light crossing time of BH horizon 
$t_{\rm lc} \sim r_g/c \sim 3.2\times10^2 \left( M/10^{7.5} M_{\odot} \right)$ s, 
where $r_g$ is the Schwarzshild radius and $M$ is the mass of central BH. 
This is in good agreement with the observed timescale, which may indicate that the central BH mass 
of PMN J0948+0022 is indeed small as was estimated by  \cite{2008ApJ...685..801Y} 
under the conical structure assumption. 
On the other hand, 
However, larger BH mass cannot be ruled out if the conical assumption
is not held and the emission region size is estimated to be larger by
the Doppler factor than the observed time scale.
We finally note that such an minute-scale 
variability has recently been detected by \cite{2013MNRAS.428.2450P}, too
and hence it is robust that PMN J0948+0022 shows this kind of rapid variability in the optical band.

High spatial resolution radio images toward the core of PMN J0948+0022
taken with VLBI revealed presence of parsec-scale jet and
the jet is directed at $\sim$30 deg \citep{2006PASJ...58..829D}.
Since the magnetic field direction is usually assumed to be
perpendicular to the PA, the observed optical PA of $\sim$15 deg
implies that the magnetic field direction at the emission site is
roughly transverse to the jet direction. 
This reminds us of well-known internal shock scenario
\citep{1991bja..book....1H} 
to explain the outburst.
If this is the case, relativistic shells emitted from central BH
intermittently with an interval of $r_g/c$ collide at $r \sim r_g
\left( \Gamma/10 \right)^2 \approx 1.5\times 10^{16}$ cm and internal 
transverse shock is generated there.  
Compression of magnetic field by the transverse shock would generate
highly ordered magnetic field inferred from the observed high PD.
The variability timescale for hour-scale suggests the size of the emission region of
$R \sim c \delta t_{\rm var} = 1 \times 10^{15} \left( \delta /10
\right) \left( t_{\rm var}/3300 \ {\rm s} \right)$ cm, which is
consistent with the standard conical jet assumption ($R \sim r
\theta$) of the opening angle of $\theta \sim 1/\Gamma$. 
We speculate that the transverse shock generated
relatively compact emission region of $R \sim 10^{15}$ cm with highly
ordered magnetic field.

If the minute-scale pulse is indeed radiated by very compact and
different blob from larger emission region responsible for the
underlying hour-scale pulse, sudden change of PA may be accompanied
because magnetic field direction inside the very compact blob does not
need to be the same as that of the larger emission region.
Of course there are some possibilities that are able to explain the
constant PA  with superposition of two emission regions,
such as the scenario with accidental coincidence of magnetic vectors
in two emission regions.
We propose other possibilities to explain the mechanism
of variability of flux and polarization other than the scenario
postulating two emission regions.
Among them, the model with a single emission region 
that internal shocks are arisen from colliding relativistic shells in
the jet also can interpret the change of variability timescale 
due to an increasing size of emission region 
as reported in \cite{2010ApJ...711..445B}.
The authors predicted a temporally asymmetric light curve and our
results support this scenario.
Our results of constant PA around 19.6 UT indicates that
the outburst might be originated in single emission region
and the change of variability timescale caused by a propagation of a
emission region in the jet.

Similar short timescale (from a few minutes to a few hours) variability of
polarization are reported from
blazar jets, such as AO 0235+164, S5 0716+714 and CTA 102
\citep{2008ApJ...672...40H, 2008PASJ...60L..37S, 2013ApJ...768L..24I}.
But it should be noted that these blazars not always shows the correlation between total flux and PD in the micro variability.
In addition, simultaneous short-term photopolarimetric observations have been performed in several blazars
\citep{2003A&A...409..857A, 2007MNRAS.381L..60C, 2011A&A...531A..38A},
and only a few blazar show the correlation between the total flux and PD.
In some blazars, the multi-zone model which have several emission regions are applied to
explain these randomly correlations between the total flux and PD in micro variability \citep[e.g., ][]{2010PASJ...62...69U,2012JSARA...7...33R}.
On the other hand, 
our results indicates the model with a single emission region
in the jet of PMN J0948+0022 as mentioned above.
This assumption might be reflecting a difference in the formation rate of blobs
between NLSy1s and blazars.
Namely, a pure synchrotron radiation from intermittent emergence of single blob can be seen in RL-NLSy1 jets. 
The low values of PD in quiescent states and no significant underlying component in MJD 56281
(PF$_C < 0.2 \times 10^{-12}$ erg cm$^{-2}$ s$^{-1}$) support this scenario.
Of course there is only one short-term photopolarimetric observation of PMN J0948+0022,
thus more photopolarimetric observations of RL-NLSy1s
in the short time variability will be needed to study the
relation between RL-NLSy1 jets and blazar jets.

We also found that the total flux increased by a factor of $\sim$2 on
MJD 56342, 61 days after the first observation (see also
Figure~\ref{fig:LC_month}). 
Motivated by this brightening, we conducted continuous polarimetric
observation over the subsequent two nights. 
As shown in Figure~\ref{fig:LC_daily1} and Figure~\ref{fig:LC_daily2}, temporal profiles of the total flux, PD and PA did
not show significant variation during the exposures, but we note that
PD was somewhat increased from $<10$\% to $\sim$15\% compared to the
previous observations on MJD 56283, 56284, and 56295. 
These results would be understood in terms of the large-scale emission
region of $R \gtrsim c \delta t_{\rm var} = 3 \times 10^{16} \left(
\delta /10 \right) \left( t_{\rm var}/10^5 \ {\rm \ s} \right)$ cm, 
by substituting the variability timescale greater than $10^5$ s.
In addition, PDs on MJD 56342 are lower than that on MJD 56281.
No significant micro variability except the data on MJD 56281 
indicates that there are various timescales of variability.
In general, the size of emission region is related with 
the location of emission region in the jet. 
These differences of emission regions give weak correlation 
between the total flux and PD on long-term variability.

In conclusion, we performed optical polarimetric observation of RL-NLSy1 
PMN~J0948+0022 using Kanata/HOWPol and Murikabushi/MITSuME 
on December 2012 to February 2013 after the near-infrared and
$\gamma$-ray flux enhancements. 
Our findings are  
(1) highly variable total and polarized fluxes but almost constant PA, 
(2) Minute-scale variability in the light curve of polarized flux, 
(3) very high maximum PD of $\sim$36\%, and 
(4) PA is grossly directed to the par sec-scale jet. 
The high and dramatic change of PD indicates that synchrotron emission
in a highly ordered magnetic field is responsible for the optical emission.
The observed variability timescale of minutes is in good agreement with the 
light crossing time of Schwarzshild radius of the central BH under the assumption of $M_{\rm BH} \sim 10^{7.5} M_{\odot}$.
We note that The transverse shock inside a jet (known as ``shock-in-jet'' model) is
consistent with the jet-directed PA and compression of magnetic field
by the shock can generate highly ordered magnetic field.
We also considered models of two radiation region scenario as well as 
that of a single emission region scenario.
Our results of constant PA around 19.6 UT indicates that the outburst might
be originated in the single emission region while other many blazar 
explained with the multi-zone model.
It might be reflecting a difference of jets
between NLSy1s and blazars if our assumptions are correct.
We also searched relation between the appearance of micro variability and the flux 
state, but did not find clear correlation.
This result indicates that the mechanism of flux variability is different between the long-term 
(from a few days to a few weeks) and the short-term component (intra-night).

This work is supported by Japan Society for the Promotion of Science, Grants-in-Aid for Scientific Research 
Nos, 17684004, 20740107, 21018007, 23340048, 14GS0211 and 19047003 and 
Optical and Near-infrared Astronomy Inter-University Cooperation Program
by the Ministry of Education, Culture, Sports, Science and Technology of Japan.

\bibliographystyle{apj}
%\bibliography{ref_PMN0948}
%%%%%%%%%%%%%%%%%%%%%%%%%%%%%%%%%%%%%%%%
%% BIBTEX OUTPUTS  %%%%%%%%%%%%%%%%%%%%%
%%%%%%%%%%%%%%%%%%%%%%%%%%%%%%%%%%%%%%%%

%%%%%%%%%%%%%%%%%%%%%%%%%%%%%%%%%%%%%%%%
%%%%%%%%%%%%%%%%%%%%%%%%%%%%%%%%%%%%%%%%
%%%%%%%%%%%%%%%%%%%%%%%%%%%%%%%%%%%%%%%%
\end{document}